\documentstyle[12pt]{article}
\newcommand{\bb}{\begin{eqnarray}}
\newcommand{\ee}{\end{eqnarray}}

\newcommand{\m}{\mu}
\newcommand{\n}{\nu}

\newcommand{\pl}{\partial}

\newcommand{\cd}{\cal {D}}
\newcommand{\spl}{\partial\hspace{-.2cm}/}
\newcommand{\psb}{\bar{\psi}}

\begin{document}
\begin{titlepage}
\begin{center}
\vspace*{.5cm}
\hspace*{10cm} TUM-HEP-232/95\\
\hspace{10cm} hep-th/9512123\\
\vspace*{1.5cm}

{\LARGE On the invariance of the string partition function under duality}\\

\vspace{1cm}

{\large Alexandros A. Kehagias}
\footnote{Supported by an Alexander von Humboldt-Stiftung}\\
\vspace{-.2cm}
{\large Physik Department \\
\vspace{-.2cm}
Technische Universit\"at M\"unchen\\
\vspace{-.2cm}
D-85748 Garching, Germany\\
E-Mail: kehagias@physik.tu-muenchen.de}\\
\end{center}
\vspace{1.3cm}

\begin{center}
{\large Abstract}
\end{center}
\vspace{.3cm}

We consider the $N\!=\!1$ supersymmetric $\sigma$-model and we examine
the transformation properties of the  partition function  under
target-space duality. Contrary to what one would expect, we find
that it is not, in general, invariant. In fact,
besides the dilaton shift emerging from the Jacobian of the duality
transformation of the bosonic part,
there also exist a Jacobian for the fermionic part since fermions are also
transform under the duality process. The latter is just the parity of the spin
structure of the word sheet and since it cannot be compensated the dual theory
is not equivalent to the original one.

\vspace{2.6cm}

\noindent
December 1995
\end{titlepage}
\newpage

Target-space duality  originally  appeared
as a discrete symmetry of the spectrum
of states of the closed string theory compactified on a torus \cite{1}.
It was later realized that the same symmetry can be extended to arbitrary
backgrounds with at least on space-like abelian isometry
 \cite{2}--\cite{dil3}.
One may  gauge this isometry and by adding a Lagrange multiplier may  render
the gauge field strength zero. The original theory is recovered by integrating
the Lagrange multiplier while by integrating over the gauge field and then
fixing the gauge the dual
 theory is obtained. The resulting background is in general
different (even topologically) from the original one
and, in addition, there exists a change
of the fermion fields \cite{2}. The original and the dual theory, although they
seem to be different, are
equivalent as classical theories. However, the measure in the partition
function is also transforms under the process. In particular, the bosonic part
of the measure give rise to the dilaton shift which is  necessary for conformal
invariance of the dual theory (if the original one is).
On the other hand, as we will see,  the fermion measure transforms
as well  under duality
 which is ultimate connected to the different way
that left and right-handed fermions are transformed.

Let us consider a $N=1$ supersymmetric $\sigma$-model defined on a 2-dim
space-time $\Sigma$ with metric ${g^{(2)}}_{\alpha\beta}$ of $(-,+)$ signature
and no antisymmetric field.
The  action for this model, in the
conformal gauge, is
\bb
S=\frac{1}{2\pi} \int d^2\sigma\left( \frac{1}{2}\pl_\m X^I\pl^\m X^Jg_{IJ}
+ig_{IJ}\bar{\psi}^I D\hspace{-.25cm}/ \psi^J+\frac{1}{6}R_{IJKL}\bar{\psi}^I
\psi^K\bar{\psi}^J\psi^L\right)
\, , \label{sa}
\ee
where the scalars $(X^I,I,J=1,\cdots,D)$ parametrize the D-dimensional
target space $M$ and the world-sheet Majorana
fermions $\psi^I$ are target-space
vectors. The metric on $M$ is $g_{IJ}$, the Riemann tensor
$R_{IJKL}$ is with respect to the Christofell connection ${\Gamma^I}_{JK}$ and
we have defined
$ D\hspace{-.25cm}/ \psi^I=\spl\psi^I+{\Gamma^I}_{KL}\spl X^K\psi^L$,
($\spl=\gamma^\m\pl_\m$ with $\gamma^\m=(i\sigma^2,\sigma^1)$ and
$\bar{\gamma}=\gamma^0\gamma^1$ is the corresponding $\gamma^5$ matrix in 2
dimensions).
The action (\ref{sa}) is invariant under the supersymmetry transformations
\bb
\delta X^I&=&\bar{\epsilon} \psi^I \, ,\nonumber\\
\delta\psi^I&=&-i\spl
X^I\epsilon-{\Gamma^I}_{KL}\bar{\epsilon}\psi^K\psi^L \, .
\label{sup}
\ee
as well as under reparametrizations (diffeomorphisms) of $M$
\bb
{X^I}^\prime={X^I}^\prime(X^J)&,& {\psi^I}^\prime=\frac{\pl {X^I}^\prime}
{\pl X^J}\psi^J \, , \label{diff}
\ee
which, moreover, commute with the supersymmetry transformations
of eq.(\ref{sup}).  The  partition function of the theory is obtained by
integrating  over all $(X^I,\psi^I)$ fields and it is given by
\bb
Z=\int d\m_bd\m_f e^{-iS}=
\int [\sqrt{g}{\cal{DX}}^I][\frac{1}{\sqrt{g}}
{\cal D}\psi^I]e^{-iS}\, ,\label{pf}
\ee
where $g=det(g_{IJ})$ is the determinant of the target-space metric.
Analytic continuation to Euclidean time is necessary for a proper definition of
the partition function in eq.(\ref{pf}). However, such a continuation can be
performed  after  the duality transformation.
We have not considered the ghost action because the ghost sector is not
affected by the presence of the background metric \cite{ghost}
and thus, it is irrelevant for our purposes. The measure $d\m_bd\m_f$ has
been defined  by  requiring that
\bb
1&=&\int d\m_f e^{-||\delta\psi||^2} \nonumber \\
1&=&\int d\m_b e^{-||\delta x||^2}\, ,
\ee
for the fermionic  $d\m_f$ and bosonic $d\m_b$ measure, respectively, where
\bb
||\delta\psi||^2&=&\frac{1}{\pi}\int d^2\sigma \sqrt{g^{(2)}}g_{IJ}
\delta\bar{\psi}^I\delta\psi^J \, ,\nonumber \\
||\delta x||^2&=&\frac{1}{\pi}
\int d^2\sigma \sqrt{g^{(2)}}g_{IJ}\delta X^I\delta X^J\, .
\ee
In this way   the measure employed in eq.(\ref{pf}) is obtained.
It should be noted also that it is, in addition, invariant under target-space
reprarametrizations (eq.(\ref{diff})). Clearly, $d\m_b$ is invariant
under eq.(\ref{diff}) and $d\m_f$ is also invariant since $\psi^I$ are
Grassmann and thus $d\psi^I$ transforms with the inverse Jacobian
(For more details  see ref.\cite{mes}).

 We will assume now, that $M$ has an abelian isometry,
orthogonal, for simplicity, to the surfaces of transitivity, so that
the target space metric may  be written as
\bb
g_{IJ}(X^K)=(g_{ij}(X^k),g_{00}(X^k))\, , \label{met}
\ee
where $X^K=(X^k,X^0=X)\, , (i,j,k=1,\cdots,d-1)$
are adapted coordinates in the target space so that the isometry is generated
by $\pl/\pl X^0$ in these coordinates.
The action of this model, as follows from eqs.(\ref{sa},\ref{met}) is then
\bb
S&=&\int d^2\sigma \left( \frac{1}{2}\pl_\m X^i\pl^\m X^jg_{ij}+\frac{1}{2}
\pl_\m X\pl^\m Xg_{00} +\frac{1}{2}ig_{ij}\psb^i(\spl\psi^j+{\Gamma^j}_{kl}
\spl X^k\psi^l+{\Gamma^j}_{00}\spl X \psi^0)+\right. \nonumber \\
& & \left.\frac{1}{2}ig_{00}\psb^0(
\spl \psi^0+{\Gamma^0}_{0k}\spl X\psi^k+{\Gamma^0}_{k0}
\spl X^k\psi^0)+\frac{2}{3}R_{0i0j}\psb^0\psi^0
\psb^i\psi^j
+\frac{1}{6}R_{ijkl}\psb^i\psi^k\psb^j\psi^l\right)\, ,\label{ss}
\ee
where the fermions have   accordingly been split as $\psi^I=(\psi^i,\psi^0)$.
The dual model of (\ref{ss}) can be constructed as usual, by employing the
first order formulation \cite{2},\cite{3} where one replaces
 $\pl_\m X$ by the gauge field (1-form) $A_\m$. Then, by  enforcing the
constraint $\epsilon^{\m\n}\pl_\m A_\n=0$, one gets  that $A_\m$
is a pure gauge, i.e. $A_\m=\pl_\m X$.  Thus,  we express
the partition function in the first order formulation as
\bb
Z=\int [\sqrt{g}{\cd A}{\cal D}\tilde{X}^0{\cd}X^i]d\m_f
e^{-iS_1-iS_2}
\, , \label{pf1}
\ee
where
\bb
S_1&=&\int d^2\sigma \left( \frac{1}{2}\pl_\m X^i\pl^\m X^jg_{ij}+
\frac{1}{2}ig_{ij}\psb^i(\spl\psi^j+{\Gamma^j}_{kl}
\spl X^k\psi^l)+\right. \nonumber \\
& & \left.\frac{1}{2}ig_{00}\psb^0(
\spl \psi^0+{\Gamma^0}_{k0}\spl X^k\psi^0)+\frac{2}{3}R_{0i0j}\psb^0\psi^0
\psb^i\psi^l
+\frac{1}{6}R_{ijkl}\psb^i\psi^k\psb^j\psi^l\right)\, ,\label{sa1}
\ee
is the $A_\m$-independent part and
\bb
S_2&=&\int d^2\sigma \left( \frac{1}{2}
A_\m A^\m g_{00} +\frac{1}{2}ig_{ij}\psb^i{\Gamma^i}_{00}A\hspace{-.25cm}/
  \psi^0+
\frac{1}{2}ig_{00}\psb^0
{\Gamma^0}_{0k}A\hspace{-.25cm}/ \psi^k-\epsilon^{\m\n}\pl_\m A_\n
\tilde{X^0}\right)\, ,
\label{s2}
\ee
is the $A_\m$-containing one. The constraint has been incorporated by means
of the Lagrange multiplier $\tilde{X}^0$. By integrating over  $\tilde{X}^0$
one gets  the original
model as usual and by integrating over $A_\m$  the dual model is obtained.
The latter
integration gives
\bb
Z=\int [\sqrt{\tilde{g}}{\cd}\tilde{X}^0{\cd}X^i]
{\cal{M}}d\m_f e^{-i\tilde{S}}
\, , \label{dpf}
\ee
where
\bb
\tilde{S}&=&\int d^2\sigma \left( \frac{1}{2}\pl_\m X^i\pl^\m X^jg_{ij}+
\frac{1}{2}\frac{1}{g_{00}}
\pl_\m \tilde{X}^0\pl^\m \tilde{X}^0
+\frac{1}{2}ig_{ij}\psb^i(\spl\psi^j+{\Gamma^j}_{kl}
\spl X^k\psi^l\right. \nonumber \\ && \left.
-\frac{1}{g_{00}}{\Gamma^i}_{00}\gamma^\m {\epsilon^\n}_\m
\pl_\n \tilde{X}^0 \psi^0)+\frac{1}{2}ig_{00}\psb^0(
\spl \psi^0-\frac{1}{{g_{00}}^2}\Gamma_{00k}\gamma^\m{\epsilon^\n}_\m
\pl_\n \tilde{X}^0\psi^k+{\Gamma^0}_{k0}
\spl X^k\psi^0)\right. \nonumber \\ && \left.
+\frac{1}{8g_{00}}\psb^i\Gamma_{i00}\gamma^\m\psi^0\psb^j
\Gamma_{j00}\gamma_\m\psi^0+
\frac{1}{8g_{00}}\psb^i\Gamma_{i00}\gamma^\m\psi^0\psb^0
\Gamma_{00k}\gamma_\m\psi^k+
\frac{1}{8g_{00}}\psb^0\Gamma_{00i}\gamma^\m\psi^i\psb^0
\Gamma_{00j}\gamma_\m\psi^j\right. \nonumber \\ && \left. +
\frac{1}{8g_{00}}\psb^0\Gamma_{00k}\gamma^\m\psi^k\psb^i
\Gamma_{i00}\gamma_\m\psi^0
+ \frac{2}{3}R_{0i0j}\psb^0\psi^0
\psb^i\psi^l
+\frac{1}{6}R_{ijkl}\psb^i\psi^k\psb^j\psi^l\right)\, ,\label{sad}
\ee
is the dual action and $\tilde{g}=\frac{1}{g_{00}^2}g$.
Note also the
factor  $\cal{M}$  in eq.(\ref{dpf}) which  emerges because
the original model has been obtained by integrating
 over scalars ($\tilde{X}^0$),
while the dual model results after  integrating over
1-forms ($A_\m$). The factor ${\cal M}$ appears then because
the number of scalars (0-forms) is in general different
from the number of 1-forms and it is explicitly given by
 \bb
{\cal{M}}=\frac{\left(\int [{\cal{D}}\tilde{\cal{X}}^0]
e^{-i\int d^2\sigma \pl_\m X^0\sqrt{g_{00}}\pl_\m X^0}\right)^2}
{\int [{\cd {A}}]e^{-i\int d^2\sigma A_\m \sqrt{g_{00}}A_\m}}
\, .
\ee
In order to evaluate ${\cal M}$ one has to analytically continue
 to Euclidean space.
In the case then of constant $g_{00}=R^2$,
$\cal{M}$ is simply $R^{B_1-2B_0}$ where
$B_0,B_1$ are the numbers of scalars and 1-forms on the surface $\Sigma$ in a
lattice regularization. By removing the lattice we are left only with the zero
mode contribution, i.e.,  $B_1-2B_0=b_1-2b_0=-\chi$ where $\chi$ is the Euler
number. Thus,  $\cal{M}=R^{-\chi}$ and it
 corresponds exactly to the  dilaton shift
$\Phi-2\ln R$ \cite{dil1}.
The non-constant $g_{00}$ case can be worked out as well.
In this case  we have
\bb
{\cal{M}}=\frac{det_0^{-1}(\sqrt{g_{00}})}
{det_1^{-1/2}(\sqrt{g_{00}})}=
\frac{exp\left(-Tr(\ln \sqrt{g_{00}}e^{-\Delta_0/M^2})\right)}
{exp\left(-\frac{1}{2}Tr(\ln \sqrt{g_{00}}e^{-\Delta_1/M^2})\right)}
\, , \label{det}
\ee
where $det_0,det_1$ denote the determinants coming from integrations over
 zero and
one-forms, respectively. We have  regularized these determinants by the
inclusion
of the $e^{-\Delta_0/M^2}, e^{-\Delta_1/M^2}$ factors where $\Delta_0,\Delta_1$
are Laplace-Beltrami operators for scalars and 1-forms, respectively.
This regularization  corresponds exactly to
the lattice regularization  mentioned above. Since
\bb
\Delta_0&=&-\nabla^2\, , \nonumber \\
\Delta_1&=&-g^{(2)}_{ab}\nabla^2+R^{(2)}_{ab}\, ,
\ee by  using the short-time expansion of the heat kernel \cite{ker}
for $\nabla^2$
\bb
Tr e^{-t\nabla^2}=\frac{1}{4\pi t}+\frac{R^{(2)}}{12\pi} +{\cal{O}}(t)
 \, , \nonumber
\ee
 we get
\bb
Tr(\ln \sqrt{g_{00}}e^{-\Delta_0/M^2}) &=& \frac{1}{2}
\int d^2x \sqrt{g^{(2)}}\ln g_{00}\left(
 \frac{M^2}{4\pi}+\frac{R^{(2)}}{12\pi}+
{\cal{O}}(M^{-2})\right)\, ,\nonumber \\
Tr(\ln \sqrt{g_{00}}e^{-\Delta_1/M^2}) &=& \frac{1}{2}
\int d^2x \sqrt{g^{(2)}}\ln g_{00} e^{-R^{(2)}/M^2}
\left( \frac{M^2}{2\pi}+\frac{R^{(2)}}{6\pi}+{\cal{O}}(M^{-2})\right)\, .
\label{mmm}\ee
Putting all together we find in the the $M\rightarrow \infty$ limit
\bb
{\cal{M}}= exp\left(-\frac{1}{8\pi}\int
d^2x\sqrt{g^{(2)}} \ln g_{00} R^{(2)}\right) \, ,
\ee
which corresponds to the well known dilaton shift \cite{dil2}. Note that there
is no infinities since the dangerous $M^2$ terms in eq.(\ref{mmm}) exactly
cancel each other.

The dual action (\ref{sad}) above is not manifest $N=1$
supersymmetric. However, it can be expressed in a manifest $N=1$ form
by the transformation
\bb
g_{00}&\rightarrow& \frac{1}{g_{00}}\, , \nonumber \\
\psi^0&\rightarrow& -\frac{1}{g_{00}}\bar{\gamma}\psi^0 \, . \label{tr}
\ee
Then, (\ref{sad}) turns out to be
\bb
\tilde{S}&=&\int d^2\sigma \left(
\frac{1}{2}\pl_\m X^i\pl^\m X^j\tilde{g}_{ij}+
\frac{1}{2}
\pl_\m \Phi\pl^\m \Phi\tilde{g}_{00} +\frac{1}{2}i\tilde{g}_{ij}
\bar{\tilde{\psi}}^i
(\spl\tilde{\psi}^j+{\tilde{\Gamma}^j}_{kl}
\spl X^k\tilde{\psi}^l+{\tilde{\Gamma}^i}_{00}\spl \Phi
\tilde{\psi}^0)+\right. \nonumber \\
& & \left.\frac{1}{2}i\tilde{g}_{00}\bar{\tilde{\psi}}^0(
\spl \tilde{\psi}^0+{\tilde{\Gamma}^0}_{0k}\spl \Phi
\tilde{\psi}^k+{\tilde{\Gamma}^0}_{k0}
\spl X^k\tilde{\psi}^0)+\frac{2}{3}\tilde{R}_{0i0j}
\bar{\tilde{\psi}}^0\tilde{\psi}^0
\bar{\tilde{\psi}}^i\tilde{\psi}^j
+\frac{1}{6}\tilde{R}_{ijkl}
\bar{\tilde{\psi}}^i\tilde{\psi}^k\bar{\tilde{\psi}}^j
\tilde{\psi}^l\right)\, ,\label{dsas}
\ee
where we have define
\bb
 \tilde{g}_{ij}=g_{ij}& , &
\tilde{g}_{00}=\frac{1}{g_{00}}\, , \nonumber \\
 \tilde{\psi}^i=\psi^i & , &
\tilde{\psi}^0=-g_{00}\bar{\gamma}\psi^0 \, . \label{def}
\ee
Thus, the bosonic parts of the original and the dual theory are related by
the well known duality transformation \cite{2}. However,
the fermion fields in the abelian direction  transforms as well
and it should be stressed that, in view of the
$\bar{\gamma}$ term,  positive and negative chirality spinors
transform differently under duality \cite{2},\cite{2''}.
In this way,  the  transformation (\ref{tr})
 converts the dual action in a manifest $N=1$ supersymmetric
form. If, moreover, the measure is invariant
under this transformation, one may
conclude that the original and the dual pictures are descriptions of the same
theory.

Ignoring the change in the bosonic measure which effectively
results in the dilaton shift, we will examine the transformation of the
fermionic measure
\bb
d\m_f=\prod_{\sigma} \frac{1}{\sqrt{g_{00}(\sigma)}}d\psi^0(\sigma)
\prod_{\sigma} \frac{1}{det(\sqrt{g_{ij}(\sigma)})}d\psi^i(\sigma)\, .
\ee
Since only $\psi^0$ is transformed, we will consider only the first part
which, expressed in positive and negative chirality
 spinors $\psi_+^0,\psi_-^0$,
respectively, is written as
\bb
\prod_{\sigma} \frac{1}{\sqrt{g_{00}}}d\psi^0
=\prod_{\sigma} \frac{1}{\sqrt{g_{00}}}d\psi_+^0
\prod_{\sigma} \frac{1}{\sqrt{g_{00}}}d\psi_-^0\, .\label{+-}
\ee
The part of the action that contains $\psi^0_+,\psi^0_-$ is
\bb
S_0=\int d^2\sigma \left(\frac{1}{2}ig_{00}\psb^0(
\spl \psi^0+2{\Gamma^0}_{0k}\spl X\psi^k+{\Gamma^0}_{k0}
\spl X^k\psi^0)+\frac{2}{3}R_{0i0j}\psb^0\psi^0
\psb^i\psi^j\right)\, , \label{so}
\ee
and by defining $\psi_\pm=\sqrt{g_{00}}\psi^0_\pm$ we
get
\bb
S_0=-\int d^2\sigma
\left(i\psb_+\pl_-\psi_++i\psi_-\pl_+\psi_-+i\psi_+\pl_-X\Gamma_k\psi_-^k
+i\psi_-\pl_+X\Gamma_k\psi_-^k
-\frac{2}{3}{R^0}_{i0j}\psi_-\psi_+
\psb^i\psi^j\right)\, , \label{so'}
\ee
where $\Gamma_k=2\pl_k\sqrt{g_{00}}$,  $\sigma^\pm=(\sigma^0\pm\sigma^1)$ and
$\pl_\pm=\pl/\pl\sigma^\pm$. The dual action is now of exactly
 the same form  with $\tilde{\psi_\pm}=
\mp\psi_\pm$ \cite{2}.
 The measure (\ref{+-}) is written simply in terms of $\psi_\pm$ as
\bb
\prod_{\sigma} \frac{1}{\sqrt{g_{00}}}d\psi^0
=\prod_{\sigma} d\psi_+d\psi_-\, ,\label{+--}
\ee
and we will examine how it transforms under
\bb
\psi_\pm\rightarrow \mp\psi_\pm \, .\label{..}
\ee
Let us choose  hermitian operators $H_+,H_-$ and let us consider the
corresponding eigenvalue problems
\bb
H_\pm\phi_n^\pm (\sigma)=\lambda_n^\pm\phi_n^\pm(\sigma)\, .
\ee
We may expand $\psi_+,\psi_-$ in the $\phi_n^\pm$ orthonormal base  as
\bb
\psi_\pm(\sigma)=\sum_{n}a_\pm^n\phi_n^\pm(\sigma) \, , \label{expan}
\ee
so that eq.(\ref{+--}) is written as
\bb
\prod_{\sigma}d\psi_+\psi_-=\prod_{n}da_+^nda_-^n\, . \label{mess}
\ee
It is not difficult to verify that  under (\ref{..}), the measure (\ref{mess})
transforms as
\bb
\prod_n da_+^n da_-^n\rightarrow det(-\bar{\gamma})\prod_n da_+^n da_-^n=
(-1)^{dimH_+} \prod_n da_+^n da_-^n
 \, , \label{c}
\ee
where $dimH_+$ is the dimension of the $H_+$ operator.
Thus, in general one should expect a change of the measure if $dimH_+$ is odd
and, correspondingly, a non-invariance of the theory under duality
transformations.
For the expansion in eq.(\ref{expan}),
one usually employs
the eigenfunctions of the quadratic part (kinetic operator)
of the fields
in the action \cite{F}--\cite{AW}. In our case, the quadratic parts of the
$\psi_\pm$  fields are simply $\nabla_\pm=-i\pl_\pm$.
If we analytically continue to
Euclidean time, we get that $\nabla_+=\pl_z$ and $\nabla_-=\nabla_+^{\dagger}=
\pl_{\bar{z}}$ which
are no more self adjoint ($z,\, \bar{z}$ are complex coordinates).
This is connected to the fact that in 2-dimensional
space of Euclidean signature there are no Majorana-Weyl spinors and left and
right-handed spinors are complex conjugate of each other (although they can
still  be treated in an independent way \cite{aa}).
 However, in this case one  may employ \cite{F}--\cite{AW}
the self-adjoint operators
\bb
H_+=-\nabla_z\nabla_{\bar{z}} \, &,& H_-=-\nabla_{\bar{z}}\nabla_z \, .
\label{hz} \ee
The expansion (\ref{expan})
 in the eigenfunctions of the operators (\ref{hz}) turns out then to be
\bb
\psi_+(\sigma)&=&\sum_{n}a_n\phi_n(\sigma), \nonumber \\
\psi_-(\sigma)&=&\sum_{n}a_n^{\dagger}\phi_n^*(\sigma) \, , \label{expan1}
\ee
so that eq.(\ref{mess}) is written as
\bb
\prod_{\sigma}d\psi_+\psi_-=\prod_{n}da_nda_n^{\dagger}\, . \label{mess1}
\ee
Then, we find that  the measure (\ref{mess1}) transforms  as
\bb
\prod_{n}da_nda_n^{\dagger}&\rightarrow&(-1)^{n_+}
\prod_{n}da_nda_n^{\dagger}
\, , \label{aaa}
\ee
where $n_+$ is the number of zero modes of $H_+$ (since the non-zero modes are
paired and do not contribute as they are always even).
 The number now of zero modes of $H_+$
is just the number of zero modes of the chiral Diral operator $\nabla_z$
($H_+$ is positive defined).
Thus, we get that the original and the dual theory are related by (\ref{tr})
and  they are not, in general  equivalent since
\bb
Z[g]=(-1)^{n_+}Z[\tilde{g}]\, . \label{final}
\ee
The  number $n_+$
of (definite chirality) zero modes of the Dirac operator
depends on the spin structure \cite{AV},\cite{H}
(it is even or odd for the even or odd
spin structures, respectively) and thus,
the dual theory also depend on the latter.
In particular, $n_+$  depends on the spin structure for surfaces with genus
$g\leq 2$ and in addition on the conformal class of the metric for $g\geq3$
\cite{H}. However,
it is a topological invariant modulo 2 and thus $(-1)^{n_+}$
is a topological invariant term and takes the values (+1),(-1) for even and odd
spin structures, respectively. In the general case with a generic metric and
antisymmetric field, the analysis is more complicated but the result is the
same. Even in this case left and right-handed fermions
transforms in a different way
under the duality process \cite{2},\cite{2''}. This in turn  leads to  a
non-trivial Jacobian for the fermionic measure in the same way that the
integration over zero and one-forms produces the Jacobian for
the bosonic measure.
A duality invariant theory can now be written down in two ways.
 Either by summing over the even ${\it only}$ spin
structures, or by considering only left-moving fermions $\psi^0_-$. It seems
that the there is no realization of  the former possibility. However, there
exist a well known string theory which is based exactly on the latter
possibility, namely, the heterotic string and thus, heterotic string is
potentially  invariant under target-space duality. Finally,
if there exist more than one  abelian isometries and dualize  an even number
of them, then since each one contributes a $(-1)^{n_+}$ factor, the net
contribution is zero and the theory is invariant.
\vspace{.6cm}

I would like to thank S. Theisen and P. Forgacs for  discussions,
K. Skenderis for correspondence and E. Kiritsis for reading
the manuscript and comments. This works supported in parts by the EC
programs SC1-CT91-0729 and SC1-CT92-0789.

\end{document}